\documentclass[a4paper,11pt]{article}
\usepackage[a4paper, total={6in, 8in}]{geometry}
\usepackage{graphicx}
\usepackage[ruled,lined]{algorithm2e}
\usepackage{array}
\usepackage{amsmath}
\usepackage{authblk}
\usepackage{multirow}
\usepackage{hyperref}
\newcolumntype{P}[1]{>{\raggedright\arraybackslash}m{#1}}

\usepackage[authoryear]{natbib}

\providecommand{\keywords}[1]
{\small\textbf{\textit{Keywords---}} #1}

\begin{document}
\title{On the inefficiency of ride-sourcing services towards urban congestion}
\author[1]{Caio Vitor Beojone}
\author[1,\footnote{Corresponding author \\ E-mails: \texttt{caio.beojone@epfl.ch} (Caio Vitor Beojone), \texttt{nikolas.geroliminis@epfl.ch} (Nikolas Geroliminis)}]{Nikolas Geroliminis}
\affil[1]{Urban Transport Systems Laboratory (LUTS), \'Ecole Polytechnique F\'ed\'erale de Lausanne (EPFL), Lausanne, CH-1015, Switzerland}
\date{}
\setcounter{Maxaffil}{0}
\renewcommand\Affilfont{\itshape\small}

\maketitle

\begin{abstract}
The advent of shared-economy and smartphones made on-demand transportation services possible, which created additional opportunities, but also more complexity to urban mobility.
Companies that offer these services are called Transportation Network Companies (TNCs) due to their internet-based nature.
Although ride-sourcing is the most notorious service TNCs provide, little is known about to what degree its operations can interfere in traffic conditions, while replacing other transportation modes, or when a large number of idle vehicles is cruising for passengers.
We experimentally analyze the efficiency of TNCs using taxi trip data from a Chinese megacity and a agent-based simulation with a trip-based MFD model for determining the speed.
We investigate the effect of expanding fleet sizes for TNCs, passengers' inclination towards sharing rides, and strategies to alleviate urban congestion.
We show that the lack of coordination of objectives between TNCs and society can create 37\% longer travel times and significant congestion.
Moreover, allowing shared rides is not capable of decreasing total distance traveled due to higher empty kilometers traveled.
Elegant parking management strategies can prevent idle vehicles from cruising without assigned passengers and lower to 7\% the impacts of the absence of coordination.
\end{abstract}


\section{Introduction}

One of the most prominent innovations seen throughout streets around the world is the ubiquitous presence of drivers using their vehicles for on-demand transportation services.
Companies use mobile applications connected through the internet to match these drivers and their passengers in real-time.
Due to the nature of their operations, these companies are called Transportation Network Companies (TNCs), but the service itself is called ride-sourcing, e-hailing, and ride-sharing, for instance \citep{rayle_etal_2016}.
Ride-sourcing services have revolutionized mobility concepts for on-demand transportation as a result of the advantages they provide, such as convenience, door-to-door rides, low fares, etc.
On-demand transportation services sound as a promising direction to improve mobility and fight car ownership.
Moreover, many TNCs offer, among the service options, shared rides (called ridesplitting).
These services try to match passengers with a reasonably similar trip within a time window.
For TNCs and drivers, this service may yield increased profits if it is capable of matching passengers and drivers efficiently.
For the passengers, this service presents a cheaper option, but they might face longer travel distances/times.
Passengers may also have almost the same advantages as those from a taxi service as door-to-door rides and no need to search for parking.
In general, these services seem to have a positive impact on economic efficiency \citep{jin_etal_2018,tachet_etal_2017}.

Naturally, all this expansion raised several concerns regarding TNCs' operations.
Oppositely to taxis, TNCs face no limitation on the fleet size that can operate, no price control, service requirements, and other legal obligations faced by the taxi industry in most cities.
Moreover, as these services base their operations on mobile applications connected to the internet, there are concerns over issues of data privacy and security \citep{jin_etal_2018}.
\citet{rogers_2017} adds other social costs, such as diminished safety and lack of professional training.
Proper planning and regulation have vital importance in the development of shared transportation for the near future \citep{narayanan_etal_2020}.
Another point of concern is the surge pricing models used, which may considerably increase the fares in moments that driver availability is insufficient \citep{schwieterman_smith_2018}.
On the other hand, surge pricing mitigates the potential chaos of a bargaining process.
Notably, it handles the spatial-temporal imbalances between driver supply and rides demand \citep{dong_etal_2018}.

Although it is not clear whether ride-sourcing is beneficial or unfavorable (or whether it does not cause anything significantly) for traffic congestion, the path to clear it is to understand how it is replacing traditional transportation modes.
In case ride-sourcing trips are directly substituting private vehicles or taxis trips then, they should have a secondary influence on congestion \citep{sfcta_2018}.
However, if ride-sourcing competes with public transportation modes (buses, trains, metro) or inducing latent demand, then the effects on congestion should be significant.
A probable scenario for \citet{tirachini_delrio_2019} and \citet{tirachini_lobo_2019} has ride-sourcing extensively substituting public transit but only inducing latent demand to a small extent.
Additionally, it might increase vehicle kilometers traveled (VKT) when vehicles cruise for passengers or when it induces latent demand \citep{vinayak_etal_2018}.
Furthermore, their market share for locomotion follows a positive trend, but there is no evidence that it can contribute efficiently towards collective mobility.
In a recent survey across TNC users in San Francisco \citep{rayle_etal_2016}, in a question ``How would you have made this trip if TNC service was not available?'', 40\% answered by taxi, 33\% by bus, and only 6\% by car.
Thus, TNC can be an attractive alternative for public transport users, and, combined with a large number of empty vehicles, it can create additional congestion problems.
Consequences of such non-cooperative interactions are called the Price of Anarchy (PoA) \citep{koutsoupias_papadimitriou_2009}, and these selfish decisions can have catastrophic consequences to urban traffic \citep{colak_lima_gonzales_2016,olmos_etal_2018,roughgarden_2005}.
For instance, reductions in demand for buses can cause imbalances making them miss their schedule, dropping their capacity because of bus bunching \citep{sirmatel_geroliminis_2018,saw_etal_2019}.
The transportation literature observes these effects for decades as the relationship between social optimum and user equilibrium \citep{vickrey_1969}.

Hence, it is imperative to understand how TNCs' operations can interfere in traffic conditions while replacing other transportation modes to seek improvements in urban mobility.
Foremost, this understanding must cover the performance of traffic and operations.
It is critical to relate the fleet size with the average speeds and service level, which are related to mobility and accessibility measures \citep{hanson_genevieve_2017,paez_etal_2012}.
Moreover, the matching process shall have a place, and thus the impact of passengers' behavior too, in a ridesplitting scenario.
Much of the literature on ride-sourcing relies on surveys \citep{rayle_etal_2016,vinayak_etal_2018,alemi_etal_2018,lavieri_bhat_2019,dong_etal_2018}, economics \citep{zha_etal_2016,he_shen_2015}, and data regressions \citep{contreras_paz_2018}.
These studies became available because of the availability of large datasets on human mobility, which enabled studies not only for ride-sourcing but for all transportation modes, such as buses \citep{bassolas_etal_2020} and taxis \citep{hamedmoghadam_etal_2019,riascos_mateos_2020}.
Even though some surveys, such as \citet{wenzel_etal_2019}, \citet{tirachini_lobo_2019}, \citet{zha_etal_2016}, link ride-sourcing services with increased traffic, they do not consider the dynamics of congestion directly nor how these services affect urban mobility and influence congestion.

However, one must note the efforts in dispatching taxis \citep{lee_etal_2004,wong_bell_2006,ramezani_nourinejad_2018,martinez_etal_2015}, and in modeling shared taxis \citep{martinez_etal_2015,santi_etal_2014,jung_etal_2016,hosni_etal_2014}.
Other efforts aim ride-sharing systems \citep{stiglic_etal_2016,nourinejad_roorda_2016,alonso_mora_etal_2017,vazifeh_etal_2018,long_etal_2018,furuhata_etal_2013,agatz_etal_2012}, pickup and delivery problems \citep{cortes_etal_2010,berbeglia_etal_2010}, and, more specifically, dial-a-ride problems \citep{molenbruch_etal_2017,masmoudi_etal_2018,ho_etal_2018,bongiovanni_etal_2019}.
Nonetheless, these problems do not correlate TNCs' fleet size to traffic conditions, nor the passengers' behavior yet.
Ignoring the effect of congestion in the operation of ride-sourcing and ridesplitting services can influence the conclusions made.
\citet{alonso_mora_etal_2017} showed that it is possible to serve the taxi demand of Manhattan, with reductions of 30\% on the current fleet.
The paper assumes that all passengers are willing to share a ride with others and that the system has perfect information about future demand.
They also did not consider the effect of congestion due to different demand conditions or the compliance of the taxi companies to decrease their fleet size.

Therefore, this paper aims to investigate the effect of expanding fleet sizes for TNCs, passengers with different willingness to share, and operational strategies over congestion conditions under a sustainable perspective.
The investigation considered a trip-based MFD traffic model integrated into an event-based simulation to tackle the dynamics of congestion.
The traffic model considers private vehicles and TNCs' vehicles.
The dynamics of the system are based on an aggregated dynamic traffic model, the network Macroscopic Fundamental Diagram (MFD) \citep{geroliminis_daganzo_2008,loder_etal_2019}, to avoid the computational burden of micro-simulation and the lack of sufficient data to properly calibrate it.
We model interactions between travelers and vehicles with an efficient matching algorithm.
It is beyond the scope of the paper, the mode-choice modeling.
Hence, we focus on the supply of rides and its participation in traffic dynamics testing several fleet sizes and willingness to share to cover a wide range of scenarios with various values of these critical variables defined externally.
Our findings show that the profit-maximizing strategies for these services can create 37\% more congestion compared to policies that minimize the impact of selfish actions, like increasing shared rides or limiting idle vehicles cruising for passengers.
Nevertheless, a higher willingness to share can minimize waiting and travel times.
This paper contributes to the literature in the following ways: i) For the best of our knowledge, this is the first work to relate the TNCs' fleet size and dynamic congestion; and ii) Investigate the effect of fleet size, willingness to share, and `empty' vehicles (number of cruising vehicles without passengers) on the system performance; iii) Develop an elegant parking management policy for cruising vehicles that can resolve congestion while serving passengers with the same quality of service.

In the remainder of the paper, Section 2 describes the methodological framework, including the simulator architecture, the real data, the matching process for passengers, and a parking-oriented strategy to decrease the circulation of empty vehicles.
Then, Section 3 presents numerical results on the effect of fleet size, willingness to share, and parking policies in the quality of service and on network congestion.
Discussion and future work are summarized in the last section.

\section{Data and Methodology}\label{sec:data_method}

\subsection{Data description}

\begin{figure}[!ht]
    \centering
    \includegraphics{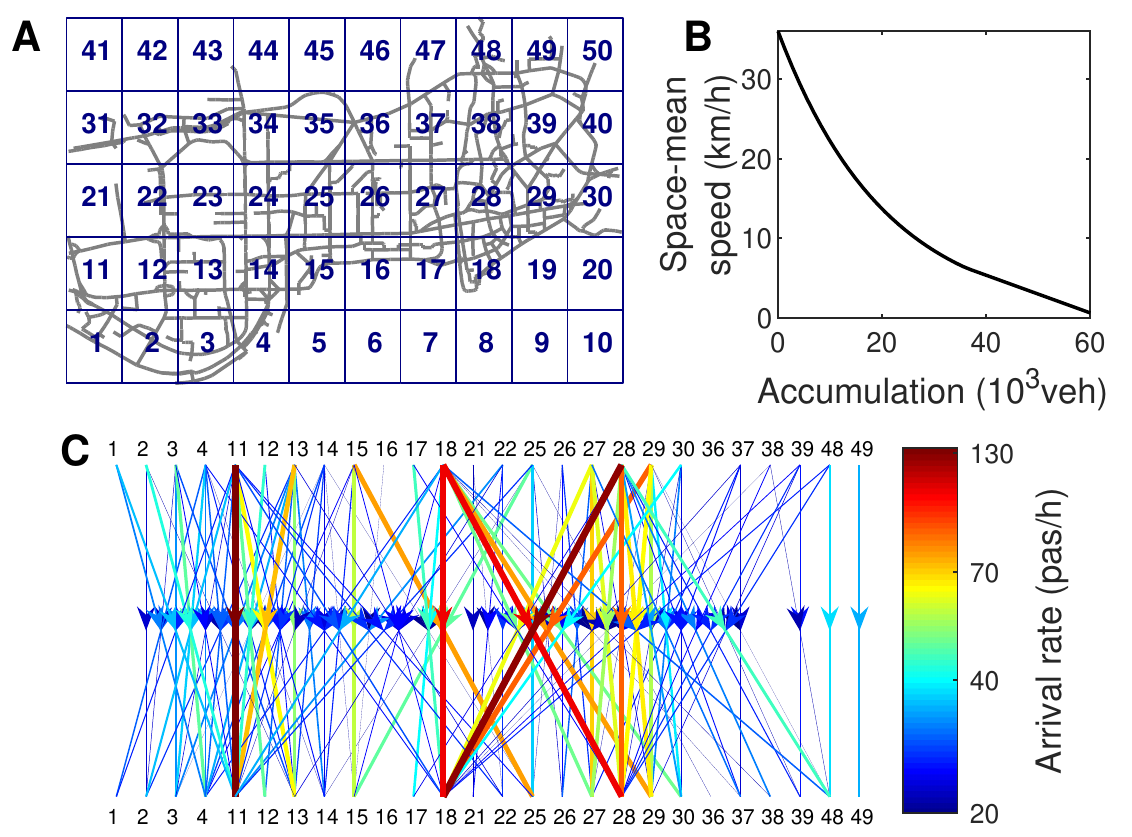}
    \caption{The simulation of a ride-sourcing service in Shenzhen used data of detailed network, demand, and traffic to provide accurate measurements on operations of ride-sourcing services. (A) Map of Shenzhen and its demand regions. (B) Space-mean speed vs Accumulation for Shenzhen. (C) Demand density per Origin-Destination pair (only arcs with demand higher than 20 requests/hour).}
    \label{fig:plot_simu_data}
\end{figure}

The original data contain GPS coordinates of 199'819 trips, with their respective origins and destinations, and 20'000 taxis every 30 seconds in the city of Shenzhen, China.
Shenzhen is immediately north of Hong Kong, in the southern province of Guangdong. Due to a fast growth period, the population was close to 11 million in 2014 \citep{ji_etal_2014}.
The development of Shenzhen came with massive foreign investments after it became a special economic zone in 1979.
The growth resulted in complex road topology and high traffic demand, leading to traffic congestion problems that propagate over time and space, creating large clusters of congested links \citep{lopez_etal_2017,bellocchi_geroliminis_2020}.
The data comprises most of the Futian and the Luohu Districts in Shenzhen, the location of the Central Business District.
The considered network consists of 1,858 intersections connected by 2,013 road segments (Fig. \ref{fig:plot_simu_data}A).
As shown in Fig. \ref{fig:plot_simu_data}A, 50 regions form the demand data.
The Origin-Destination (OD) data for the 50 regions contained 199,819 requests divided into 20 hours of the day.
Fig. \ref{fig:plot_simu_data}C shows the demand as arrows connecting origins and destinations.
The colormap represents the average arrival rate for each origin-destination arrow.
Demand is heterogenous over both time and space.
Fig. \ref{fig:plot_simu_data}B shows the assumed MFD of the region which is explained later this section. 

An MFD represents the traffic congestion and computes the average speeds in the network as a function of the accumulation of PVs and RSVs.
The MFD used on Shenzhen is based on the one obtained in \citet{ji_etal_2014} for the same data of taxi trips.
To approximate the jam accumulation for all moving vehicles in the network, we used the total road length (both ways, in case of multiple lanes) using OpenStreetMap data.
We assumed that congestion is homogeneous in the region.
Hence, a single MFD is capable of measuring congestion.
Another reason for such simplifying assumption is the computation of shortest paths – which remain unaltered during the simulation – and, consequently, the route choice.
Eq. [\ref{Eq:MFD}] shows the Accumulation n \emph{vs}. speed v(n) relationship and Fig. \ref{fig:plot_simu_data}B is the graphical representation.
While this simplification created an elegant model with small computational effort, it can still well represent the distribution of trip lengths as in real settings (Section \ref{sec:comp_results} provides more details).

\begin{align}\label{Eq:MFD}
    v(n)=\begin{cases}
            36e^{(\frac{29}{600}m)}, & \mbox{if $m\leq36$}\\
            6.31-0.28(m-36), & \mbox{if $36<m\leq60$}\\
            0, & \mbox{if $m>60$}
         \end{cases}, && \mbox{where $m \equiv \frac{n}{1000}$}
\end{align}

\subsection{State description, Congestion dynamics and matching process}

Four different entity classes populate the simulation environment: private vehicles (PVs), waiting passengers (WPs), traveling passengers (TPs), and ride-sourcing vehicles (RSVs).
Each of the classes has properties to define them shown in Table \ref{tab:nomenclature}.

\begin{table}[!ht]
\centering
\caption{Nomenclature of tuple elements for each entity.}\label{tab:nomenclature}
\begin{tabular}{lll}
\hline
Entity & Property & Description\\
\hline
\multirow{6}{1.5cm}{PV (Private Vehicle)} & $\mbox{\normalfont{PV}}^{id}_i$ & Identification\\[2pt]
 & $\mbox{\normalfont{PV}}^{at}_i$ & Arrival time\\[2pt]
 & $\mbox{\normalfont{PV}}^o_i$ & Origin\\[2pt]
 & $\mbox{\normalfont{PV}}^d_i$ & Destination\\[2pt]
 & $\mbox{\normalfont{PV}}^{rd}_i$ & Remaining distance\\[2pt]
\hline
\multirow{7}{1.5cm}{WP (Waiting Passenger)} & $\mbox{\normalfont{WP}}^{id}_j$ & Identification\\[2pt]
 & $\mbox{\normalfont{WP}}^{at}_j$ & Arrival time\\[2pt]
 & $\mbox{\normalfont{WP}}^o_j$ & Origin\\[2pt]
 & $\mbox{\normalfont{WP}}^d_j$ & Destination\\[2pt]
 & $\mbox{\normalfont{WP}}^{wts}_j$ & Willingness to share\\[2pt]
 & $\mbox{\normalfont{WP}}^{dr}_j$ & Assigned driver ID\\[2pt]
\hline
\multirow{8}{1.5cm}{TP (Traveling Passenger)} & $\mbox{\normalfont{TP}}^{id}_j$ & Identification\\[2pt]
 & $\mbox{\normalfont{TP}}^{pt}_j$ & Pick-up time\\[2pt]
 & $\mbox{\normalfont{TP}}^{o}_j$ & Origin\\[2pt]
 & $\mbox{\normalfont{TP}}^{d}_j$ & Destination\\[2pt]
 & $\mbox{\normalfont{TP}}^{wts}_j$ & Willingness to share\\[2pt]
 & $\mbox{\normalfont{TP}}^{dr}_j$ & Assigned driver ID\\[2pt]
 & $\mbox{\normalfont{TP}}^{td}_j$ & Distance traveled\\[2pt]
\hline
\multirow{8}{1.5cm}{RSV (Ride-Sourcing Vehicle)} & $\mbox{\normalfont{RSV}}^{id}_k$ & Identification\\[2pt]
 & $\mbox{\normalfont{RSV}}^{l}_k$ & Last passed node\\[2pt]
 & $\mbox{\normalfont{RSV}}^{cd}_k$ & Current destination\\[2pt]
 & $\mbox{\normalfont{RSV}}^{rd}_k$ & Remaining distance to the current destination\\[2pt]
 & $\mbox{\normalfont{RSV}}^{np}_k$ & Number of passengers inside the vehicle\\[2pt]
 & $\mbox{\normalfont{RSV}}^{pID}_k$ & ID of assigned passengers (in order of activity)\\[2pt]
 & $\mbox{\normalfont{RSV}}^{pAC}_k$ & List of activities (in order of execution)\\[2pt]
\hline
\end{tabular}
\end{table}

Vehicles move in the network following a trip-based model with an accumulation \emph{vs} speed MFD \citep{arnott_2013}.
For every new trip (by a PV or a RSV, with a passenger or empty), the model computes its total distance to the destination and updates the remaining distance for each vehicle based on \citet{lamotte_etal_2018}.
One way to introduce it starts from the simple observation that a vehicle with trip length $l_0$, which entered at time $t_0$, should exit after traveling $l_0$, i.e., after a time interval $\tau_0$ satisfying Eq.[\ref{Eq:l_0}].

\begin{align}
    l_0 = \int_{t_0}^{t_0+\tau_0} v(n(k)) \, d\!k \label{Eq:l_0}
\end{align}

The population of PVs fluctuates as every PV has a specific arrival time.
Furthermore, we assume that once a PV reaches its destination ($\mbox{PV}^{rd}_i=0$), it enters a garage or parking lot, leaving the system.
Note that RSVs and PVs move in the network at every time step at variable speeds, varying according to the traffic conditions summarized in the MFD (Fig. \ref{fig:plot_simu_data}B).

WPs are the entities that were not served yet by a RSV.
If a WP $j$ is willing to share his ride (hires the ridesplitting service), his willingness to share $\mbox{WP}^{wts}_j$ is set to 1.
Otherwise, it is set to $\mbox{WP}^{wts}_j=0$.
The choice for sharing is the result of a single Bernoulli trial for each traveler generated in the system.
Nevertheless, even if a user is willing to share, a good quality of shared service should be provided for him/her to accept such an option, otherwise s/he will travel alone.
This is described in more details later this section.
Finally, once s/he has an assigned RSV to pick-up, it cannot change, and it is linked to the passenger $j$ through $\mbox{WP}^{dr}_j$.

Once a RSV picks-up a WP, the last becomes a TP (leaves the list of WPs and adds a new member to the list of TPs).
The new TP inherits most of the data from the WP, said: identification, origin, destinations, willingness to share, and assigned driver.
However, TPs have new properties, said: time of pick-up, and distance traveled.
As the speeds might vary a lot during a trip, a traveling passenger has no information about the delivery time, which is informed by the RSV once it reaches the passenger's destination.

The central entity of the ride-sourcing service is the RSV, which is responsible for the pick-up and delivery of passengers according to their preferences.
Different from the PVs, RSVs have their positioning tracked all the simulation long.
They also may assume different states depending on their current activity.
Every RSV has an identification $\mbox{RSV}^{id}_k$, a last passed node $\mbox{RSV}^l_k$ (node in the network, updated at every time step), a current destination $\mbox{RSV}^{cd}_k$ (node in the network, such as a WP's origin, or a WP's destination, or a parking lot), and a remaining distance to the current destination $\mbox{RSV}^{rd}_k$, to keep track of their position.
To keep track of their activities, they have the destinations' ID $\mbox{RSV}^{pID}_k$ (identification of the passenger – waiting or traveling one), and number of passengers inside the vehicle $\mbox{RSV}^{np}_k$.
RSVs have a maximum capacity of two passengers.
\citet{li_etal_2019} found that only 6-7\% of trips were shared and more than 90\% of them had at most two passengers in a study in Chengdu, China, meaning that shared trips with 3 or more passengers is in the range of 0.7\% or even less.

\begin{figure}[!ht]
    \centering
    \includegraphics[width=\textwidth]{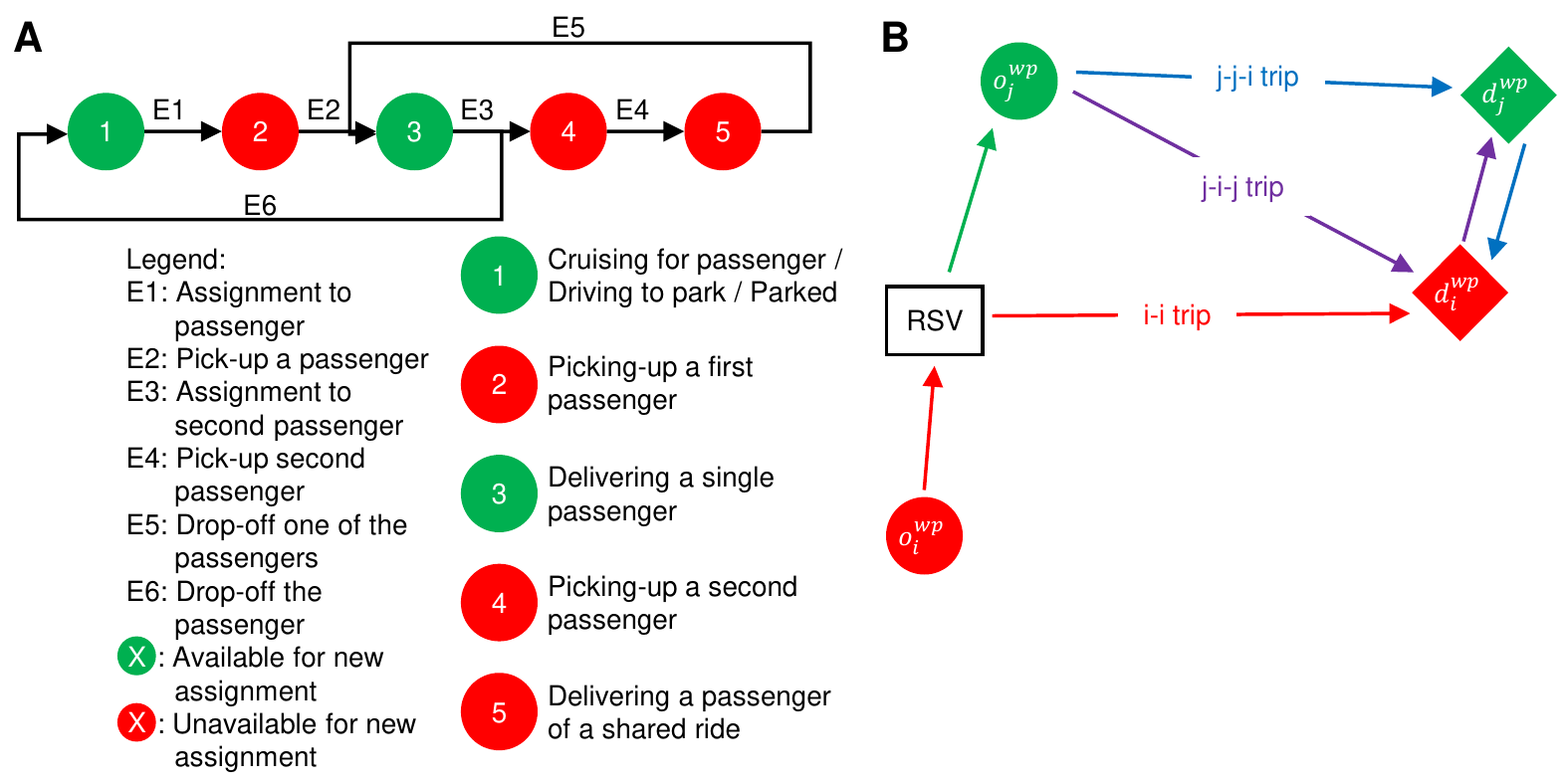}
    \caption{(A) RSV activity flow framework. (B) Ride-hailing and ridesplitting trip scheme.
    `\textit{i-i} trip' refers to a direct trip from $\mbox{\normalfont{TP}}^{o}_i$ to $\mbox{\normalfont{TP}}^{d}_i$.
    `\textit{j-j-i} trip' refers to a ridesplitting trip that will deliver passenger \textit{j} first, and then passenger \textit{i}.
    `\textit{j-i-j} trip' refers to a ridesplitting trip that will deliver passenger \textit{i}, and then passenger \textit{j}.}\label{fig:states_match}
\end{figure}

RSVs perform different activities as they move through the network.
Fig. \ref{fig:states_match}A shows how RSVs change their states during the simulation.
The states refer to the current activity of the RSV.
A RSV can perform six different activities:
\begin{itemize}
    \item Cruising for passenger: the vehicle has no passengers and is driving around his current location looking for a passenger (WP);
    \item Driving to park: the vehicle has no passengers and is driving to a parking lot near high demand areas (only possible when the parking strategy from Section \ref{sec:parking} is active);
    \item Parked: the vehicle has no passenger, reached a parking lot near high demand areas, and waits there for the next assignment (only possible when the parking strategy from Section \ref{sec:parking} is active);
    \item Picking-up a first passenger: the vehicle received the location of a waiting passenger (assignment), and it is moving towards the passenger's pick-up position (origin);
    \item Delivering a single passenger: after picking-up the passenger, the vehicle drives him/her towards the final destination;
    \item Picking-up a second passenger: the vehicle has one passenger and is moving towards a second passenger that matched the current ride; and
    \item Delivering a passenger of a shared ride: the vehicle has two passengers and is moving towards the destination of one of them.
\end{itemize}

For each activity of a RSV, the path choice follows the Floyd-Warshall algorithm (the shortest path) in terms of distance.
When evaluating a ridesplitting match (a vehicle with one passenger matches with a second passenger), two types of trip schemes arise.
Fig. \ref{fig:states_match}B illustrates both types of trips (\textit{j-i-j} and \textit{j-j-i} sequences) and direct route (\textit{i-i} sequence).

Defining a trip determines the order of the points which the vehicle will visit.
The matching process is responsible for determining RSVs' trips.
WPs have 1 minute to get an assignment (a RSV available to pick-him/her-up) that fulfills the requirements of waiting time and detour.
After this time, travelers leave the WP list and choose a new mode of transportation (busses, bike, walk, taxis, private vehicle) in an event called abandonment.
Busses, bike, and walking are considered secondary to the accumulation, and, therefore, the simulation does not keep track of their activities.
They transfer with a fixed travel time to their destination.
In the case of an abandoning passenger who decides to travel by taxi, it is modeled similarly to PVs.
While a mode-choice module could integrate the simulation, this is beyond the scope of the paper that focuses on the supply side.
The interest of this work is to analyze the effect of ride-sourcing services in congestion for different fleet sizes and willingness to share.
For demand-oriented work, the reader could refer to \citet{tirachini_lobo_2019}, \citet{tirachini_delrio_2019}, and \citet{zha_etal_2016}.
The effect of mode choice and socioeconomic characteristics in the ride-sourcing literature is a research priority.

The matching process assigns the closest RSV that fulfills all requirements to perform a ride.
It means that even a RSV that fulfills all requirements (capable) and may save some more VKT (Vehicle Kilometers Traveled) will not get the passenger if there is another capable RSV closer, for instance.

Requirements for matching passengers and drivers derive from passengers' tolerances towards deviating from their original path (where applicable) and waiting.
In all cases, the RSV must be able to reach the passenger in less than $\Delta$ minutes under current traffic conditions (\ref{Eq:match_1}).
Any idle vehicle that fulfills the later is capable of serving ride-hailing and ridesplitting passengers.
However, ridesplitting rides (actually shared rides) must fulfill additional conditions.
Firstly, all involved passengers must hired ridesplitting rides ($\mbox{WP}^{wts}_j=\mbox{TP}^{wts}_i=1$).
In the `\textit{j-i-j}' trip from Fig. \ref{fig:states_match}B, it is not allowed to add more than a maximum relative detour $\Omega$ to the trip distance of TP `\textit{i}'.
Thus, the detour of picking-up the WP `\textit{j}' must be acceptable for `\textit{i}' (Eq. [\ref{Eq:match_2}]); the same applies to `\textit{j}' regarding the delivery of `\textit{i}' (Eq. [\ref{Eq:match_3}]).
Finally, for the sequence `\textit{j-j-i}', the detour of picking-up and delivering `\textit{j}' must be acceptable for `\textit{i}' (Eq. [\ref{Eq:match_4}]).
Note that, in this sequence, there is no detour for `\textit{j}'.
The matching process considers the distance between two points ($p(\cdot,\cdot)$), and current speed ($v(t_{clock})$) to compute the conditions of Eqs. [\ref{Eq:match_1}--\ref{Eq:match_4}].
In case both sequences (`\textit{j-i-j}' and `\textit{j-j-i}') are possible, the shortest one in distance is chosen.

\noindent
\begin{align}
    p(\,\mbox{RSV}^l_k\,, \mbox{WP}^o_j\,) & \leq v(t_{clock}) \cdot \Delta \label{Eq:match_1}\\
    \mbox{TP}^{td}_i + p(\,\mbox{RSV}^l_k\,,\mbox{WP}^o_j\,) + p(\,\mbox{WP}^o_j\,,\mbox{TP}^d_i\,) & \leq p(\,\mbox{TP}^o_i\,,\mbox{TP}^d_i\,) \cdot (1 + \Omega) \label{Eq:match_2}\\
    p(\,\mbox{WP}^o_j\,,\mbox{TP}^d_i\,) + p(\,\mbox{TP}^d_i\,,\mbox{WP}^d_j\,) & \leq p(\,\mbox{WP}^o_j\,,\mbox{WP}^d_j\,) \cdot (1 + \Omega) \label{Eq:match_3}\\
    \mbox{TP}^{td}_i + p(\,\mbox{RSV}^l_k\,,\mbox{WP}^o_j\,) + p(\,\mbox{WP}^o_j\,,\mbox{WP}^d_j\,) + p(\,\mbox{WP}^d_j\,,\mbox{TP}^d_i\,) & \leq p(\,\mbox{TP}^o_i\,,\mbox{TP}^d_i\,) \cdot (1 + \Omega) \label{Eq:match_4}
\end{align}

Note that the matching requirements share many similarities with the shareability networks presented in \citet{santi_etal_2014}.
Moreover, our matching process occurs online, as seen in \citet{alonso_mora_etal_2017}.
However, these processes share some key differences:
1) we do not allow to change the vehicle that will pick-up a passenger; and
2) fully occupied vehicles are not options for assignments.
Reader can refer to \citet{martinez_etal_2015}, \citet{jung_etal_2016}, \citet{hosni_etal_2014}, \citet{stiglic_etal_2016}, \citet{nourinejad_roorda_2016}, \citet{long_etal_2018}, \citet{furuhata_etal_2013}, and \citet{agatz_etal_2012} for other matching strategies for on-demand transportation services.

\subsection{Parking strategy}\label{sec:parking}

TNCs' attractiveness depends significantly on the fast response on picking-up passengers when a request arrives (similar to other types of response systems, see for example a vast literature for emergency response systems, based on location theory).
To succeed in this objective and attract higher demand from other modes of transport, TNCs try to increase the number of registered drivers (see an economic analysis for a static model in \citet{tirachini_lobo_2019}).
While an increased fleet size could decrease the waiting time for passenger pick-up, it creates a mass of idle circulating vehicles.
As the numerical study of Section \ref{sec:comp_results} shows, strong congestion effects might appear in the network.
Nevertheless, this congestion also influences the other modes of transport moving in the same part of the network.
Thus, if there is no intervention from the government to penalize the negative externalities of these actions (e.g.\ through pricing or creating additional opportunities for public transport), TNCs can have advantage over other modes of transport.
If there is spare parking capacity, the development of simple strategies could decrease the circulation of idle vehicles without significantly increasing the waiting time.
The purpose of such is that by leaving on the side demand interactions and mode choice, we can still show with a dynamic model of congestion that smart parking strategies can have a positive effect on the overall system.
Whereas the implementation and pricing of such systems can influence mode interactions, this is yet another research direction that has to be further analyzed in transport economics terms.
This paper focuses more on the supply interactions and dynamics of congestion as a function of fleet sizes and parking strategies that try to decrease the number of circulating idle TNC vehicles.

Idle RSVs cruise, in a random walk, near their last destination waiting for their next assignment.
Such behavior has the potential to increase empty kilometers traveled and degrade traffic conditions.
For this reason, we propose an elegant parking strategy where we assign idle vehicles to parking lots near high demand areas.
The added value of parking lots is that parked vehicles do not contribute to the MFD accumulation and, thus, congestion levels are lower.
RSVs are available for new assignments while in `driving to park' or `parked'.

\begin{figure}[!ht]
    \centering
    \includegraphics{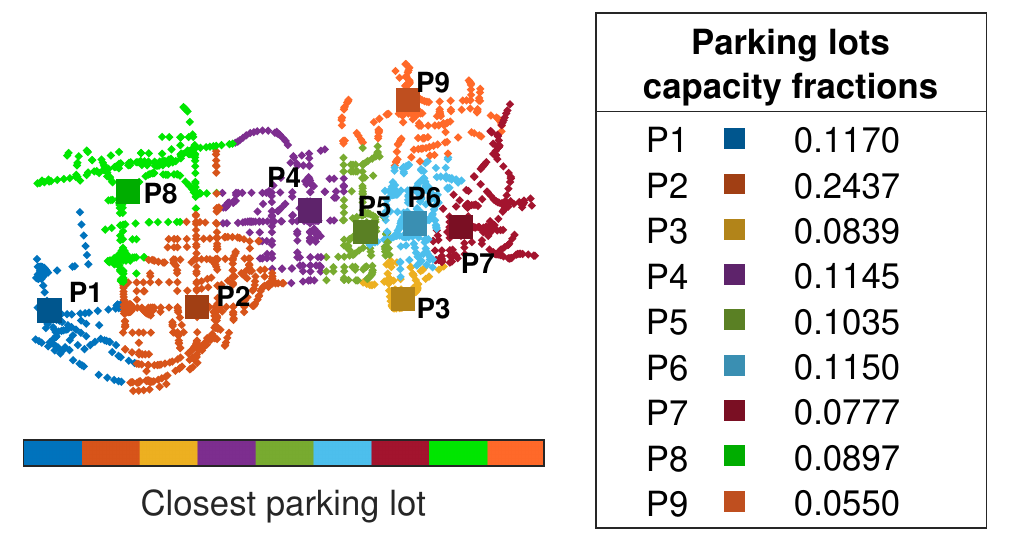}
    \caption{Location of parking lots, their closest intersections and their capacity fractions.}\label{fig:park_data}
\end{figure}

The locations of parking lots are the result of a simplified p-median problem \citep{owen_daskin_1998} in two stages.
The first stage defines the intersections with the shortest average distances to other nodes in their respective demand regions (Fig. \ref{fig:plot_simu_data}A).
The second stage solves the p-median problem for the defined nodes and the demand values for each region.
\ref{fig:park_data} summarizes the parking lots positions and their closest intersections.
We show the capacity (number of available spots) of each parking lot as the fraction of the demand it supplies in the p-median problem.
The total capacity of parking lots is equal to the fleet size of RSVs such that no vehicles are allowed to cruise for passengers.

Assignment of idle RSVs for parking lots uses a color system to prioritize emptier parking lots.
The color system consists of setting different dispatch priorities for parking lots according to their classifications (colors).
Colors are a reference to their usage level, i.e., a parking lot with fewer vehicles have a higher priority to get an assigned vehicle.
The parking strategy considers the proximity between RSV and parking lot as a secondary classification.
The Drum-Buffer-Rope method inspired this process \citep{cox_schleier_2010}.
Green parking lots have more than 70\% of available spots, yellow ones have more than 30\%, and red ones are almost full.
Algorithm \ref{alg:park} summarizes the elegant parking strategy, where $\mathcal{P}$ is the set of parking lots defined as the tuple $\mbox{PK}_p=\{\mbox{PK}^l_p,\mbox{PK}^c_p,\mbox{PK}^s_p,\mbox{PK}^a_p\}$ of the locations, capacity, number of available spots, and number of assigned vehicles, respectively.
The dynamics of vehicle occupancy in parking lots during the simulation are provided in the next section.

\begin{algorithm}[!ht]
    \KwIn{$\mbox{RSV}_k$, $\mathcal{P}$\tcp*[f]{Observed RSV and parking lots.}}
    \BlankLine
    \BlankLine
    $\mbox{ord}\mathcal{P}\gets\mbox{sort}(\mathcal{P},\mbox{RSV}_k^l)$\tcp*[r]{Sort parking lots to their proximity to the RSV.}
    $\mbox{g}\mathcal{P}\gets\mbox{ord}\mathcal{P}(^{\mbox{PK}^s_p}/_{\mbox{PK}^c_p}\!<0.3)$\tcp*[r]{Identify parking lots with a `green' flag.}
    $\mbox{r}\mathcal{P}\gets\mbox{ord}\mathcal{P}(^{\mbox{PK}^s_p}/_{\mbox{PK}^c_p}\!>0.7)$\tcp*[r]{Identify parking lots with a `red' flag.}
    $\mbox{y}\mathcal{P}\gets\mbox{ord}\mathcal{P}-\mbox{r}\mathcal{P}-\mbox{g}\mathcal{P}$ \tcp*[r]{Identify parking lots with a `yellow' flag.}
    $s\mathcal{P}\gets\mbox{concatenate}(\mbox{g}\mathcal{P}, \mbox{y}\mathcal{P}, \mbox{r}\mathcal{P})$\tcp*[r]{Regroup parking lots ordered by flags and distances.}
    $\mbox{RSV}^{cd}_k \gets \mbox{PK}^l_{s\mathcal{P}(1)}$\tcp*[r]{Update RSV's current destination.}
    $\mbox{PK}^a_{s\mathcal{P}(1)} \gets \mbox{PK}^a_{s\mathcal{P}(1)}+1$\tcp*[r]{Update the number of assigned vehicles of the parking lot.}
    \BlankLine
    \Return{$\mbox{\normalfont RSV}_k$, $\mathcal{P}$}
    \caption{Pseudo-code for parking assignment.} \label{alg:park}
\end{algorithm}

\section{Computational results}\label{sec:comp_results}

We analyzed several metrics, with different ride-sourcing fleet sizes (from 1000 to 7000 in increments of 500 vehicles -- based on the operating number of taxis in \citet{ji_etal_2014} for Shenzhen), and willingness to share (fraction of passengers hiring ridesplitting: 0\%, 30\%, 60\%, and 90\%).
The maximum waiting time, $\Delta$, and the maximum detour, $\Omega$, were set to 10 minutes and 20\% of the trip length (shortest path from origin to destination), respectively.
From all abandonments, about half choose to travel by busses, bike, or walk.
The other half call a taxi or pick-up a private vehicle (see \citet{rayle_etal_2016}).
As the number of abandonment trips is small for fleet sizes above 2000 vehicles (1-4\% from all trips, including private vehicles and ride-sourcing), variations in the fraction of these trips between public and private modes do not influence the numerical results and conclusions remain unchanged.
Scenarios, where the elegant parking strategy is active, are separated from those where it is deactivated.

A Poisson process describes the arrival process of both PVs and WPs.
They had piece-wise constant rates in a 3-hour long simulation with a low-high-low demand profile lasting one hour for each period.
In the low-demand profile, private vehicles and ride-sourcing passengers split in 34'000 and 6'000 trips per hour for each, respectively.
In the high-demand profile, arrival rates double.

Our results show that encouraging ridesplitting is not enough to decrease the VKT, a measure associated with worse congestion, fuel consumption, and safety issues.
Furthermore, traffic congestion worsens as ride-sourcing fleets grow.
Finally, the findings acknowledge that it is necessary to restrain idle ride-sourcing vehicles from cruising to decrease VKT.

As this is a trip based simulation, with an MFD representation of speed dynamics in the network, it is necessary to test whether this parsimonious model, without link speed variations and detailed traffic assignment, produces realistic traffic characteristics.
To do so, we compare trip length distributions from the real taxi trips in Shenzhen with the ones produced by the simulator.

\begin{figure}[!ht]
    \centering
    \includegraphics[width=\textwidth]{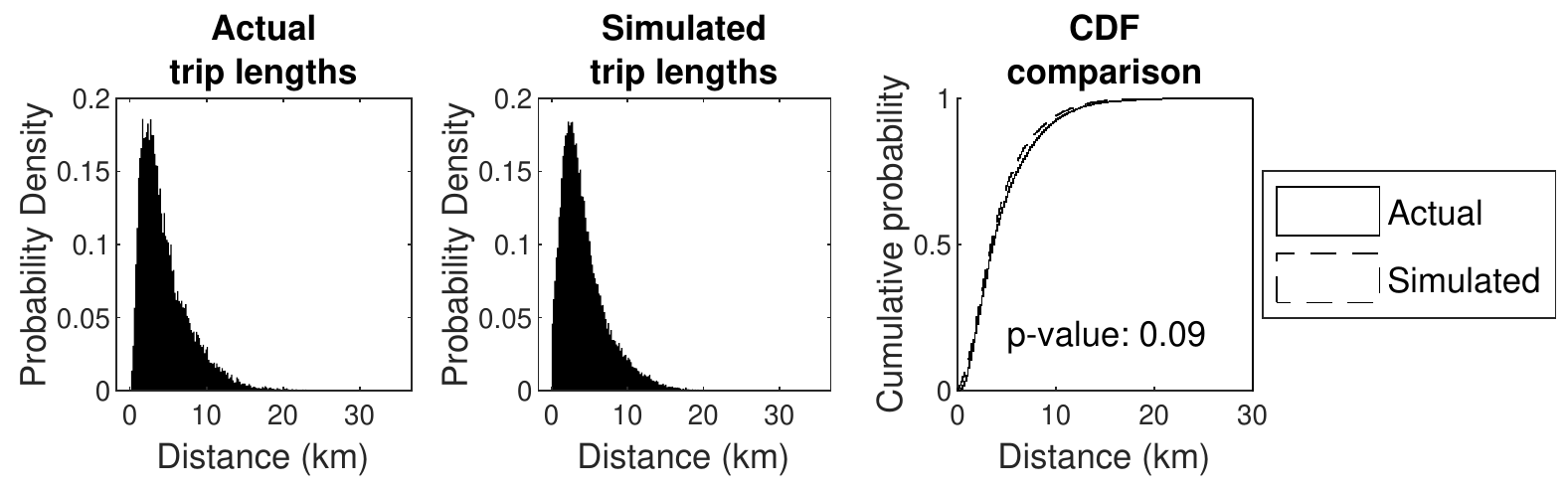}
    \caption{Histograms and CDF of trip lengths. Result of a Kolmogorov-Smirnov test comparing both samples.}
    \label{fig:hist_trip_lengths}
\end{figure}

We sampled 2'000 trip lengths between 7:00 and 10:00 from the taxi trips with passengers from Shenzhen and 2'000 trips trips from an instance of the simulation, to ensure that the simulation could provide a realistic representation.
Fig. \ref{fig:hist_trip_lengths} summarizes the probability density functions for both samples and compares their cumulative density functions.
Graphically, both samples have similar shapes.
Furthermore, a Kolmogorov-Smirnov test evaluated the similarity between the samples and did not reject the null hypothesis for a confidence level of 5\%.
The use of the shortest path and aggregating demand data in regions (see Fig. \ref{fig:plot_simu_data}A) did not generate a significant distinction between samples such that the simulator could represent trip lengths accurately.

\subsection{The effect of willingness to share and fleet size on network congestion}

Evaluating the ride-sourcing service requires a multi-dimensional look.
Performance measurements include waiting times, journey duration, and abandonments regarding the perspective of passengers.
While the fraction of travelers that are willing to share a trip is an input to the simulator (ranging from 0 to 90\%), the quality of the matching could influence the actual number of travelers that share a trip.
It depends on the extra detour for both travelers that potentially match, as described in Section \ref{sec:data_method}.

We consider that a complete journey of a passenger starts the moment s/he orders the service and ends the moment s/he reaches his destination.
However, passengers that abandon the ride-sourcing service may lead to unrealistic results.
Furthermore, waiting and traveling times may underestimate the consequences of abandonments since served trips will concentrate near main demand centers, whereas those far from them will abandon unserved.
For instance, abandonments range between 15\% and 33\% (for willingness to share of 90\% and 0\%, respectively) for a fleet size of 1500 ride-sourcing vehicles and decrease to negligible values for larger fleets.
For this reason, abandonments penalize the measurements proportionally to the ratio of abandoned passengers using Eqs. [\ref{wait}] and [\ref{journey}].

\begin{align}
    W_{i}^{'} & = W_i \times (1 + f_{ab})\label{wait}\\
    T_{i}^{j} & = (W_{i}+T_{i}^{tr}) \times (1+f_{ab})\label{journey}
\end{align}

\noindent
Here, $W_{i}^{'}$ represents the waiting time for a passenger $i$ subject to a penalty and $T_i^j$ describes the journey duration calculated from the waiting time ($W_i$), the travel time ($T_i^{tr}$) and the abandonments (as a ratio $f_{ab}$ ranging between 0 and 100\%).

\begin{figure}[!ht]
    \centering
    \includegraphics{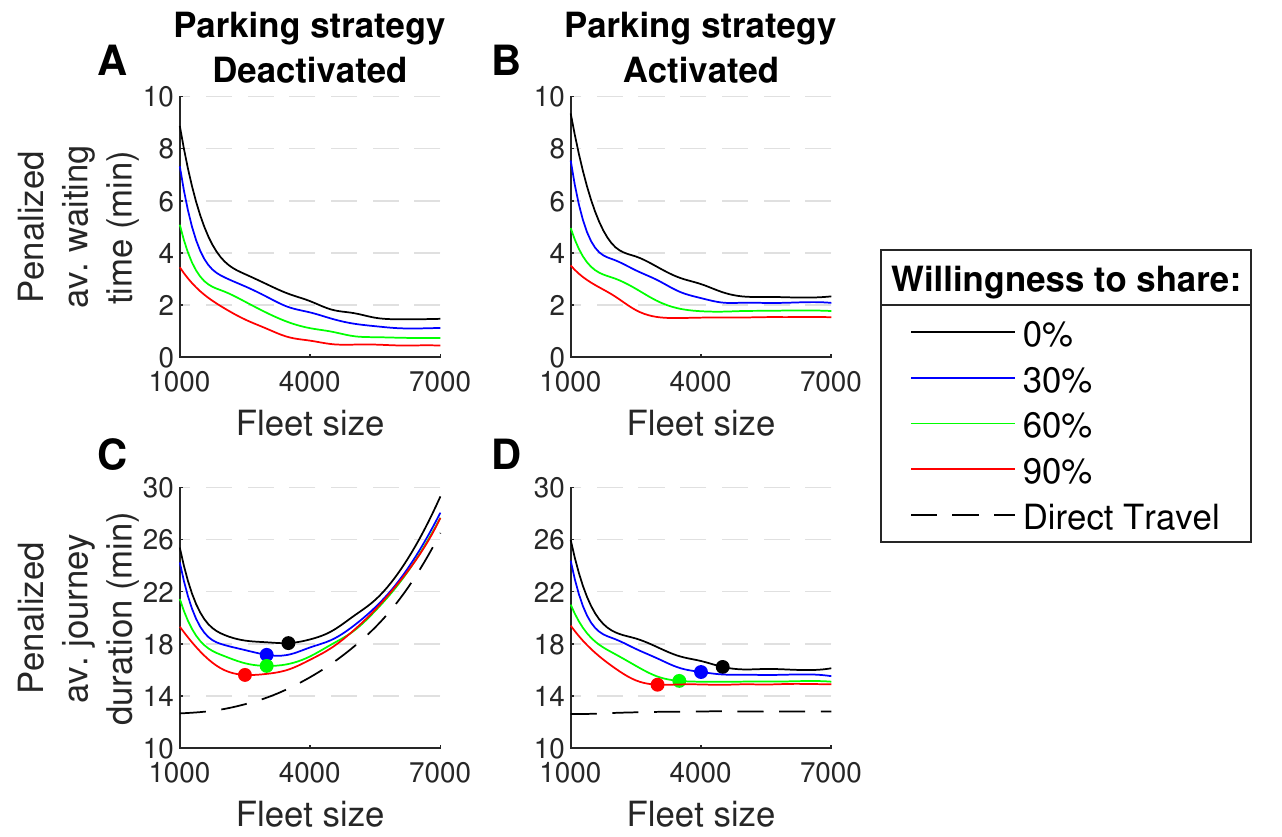}
    \caption{Optimum fleet sizes minimize the average journey duration of passengers. (A, B) Series of average waiting times for different fleet sizes and willingnesses to share. (C, D) Series of average journey duration. A and C show results for instances without use of the parking strategy. B and D show results for instances using the parking strategy. Markers indicate the fleet size with minimum journey duration. The results were corrected with a penalty to abandonments. Direct travel time represents the in-vehicle time for a direct service with no detour.}
    \label{fig:plot_wait_trip}
\end{figure}

As mentioned, the quality of service of transportation services (TNCs, taxis, metros or buses) significantly depends on the passengers' waiting times, which is close to zero for private cars.
Bringing passengers into the service requires planning on the business model, comprising fleet sizes, service availability, fares, so on.
An on-demand transportation service, such as a ride-sourcing service, has to manage the dispatching process of its fleet in real-time, accounting to the route choice, and chances to match passengers.
Operators may reposition drivers establishing fares dynamically through the city.
Once dispatching and repositioning policies are well defined and operational, decreasing the waiting times of passengers requires increases in the fleet sizes unavoidably.
For example, a passenger may wait between 3.5 and 9 minutes when only 1000 vehicles are operating (Fig. \ref{fig:plot_wait_trip}A).
However, for larger fleet sizes, a passenger may wait between 0.4 and 1.5 minutes, on average.
Such shorter waiting times can make ride-sourcing services more appealing compared to public transport \cite{hensher_rose_2007}, as the fleet size grows.
As the fleet size approaches its ideal value, the number of idle vehicles at peak-hour comes to near zero.
Any further increase in the fleet size slows the system's recovery from the peak-hour because of more extensive congestion and, thus, lower traveling speeds.

After picking up a passenger, the ride-sourcing service must plan and execute his/her delivery.
Planning includes setting a proper path/route according to a specific strategy, such as minimize travel time for the driver.
After operationalizing a route planning process, traveling times would only decrease upon higher movement speeds.
There resides the conflict of managing fleet sizes for ride-sourcing services since traveling speeds depend on traffic conditions.
Increasing fleet sizes significantly influence average travel speeds.
For instance, Fig. \ref{fig:plot_wait_trip}C presents the result of combining shorter waiting times and longer traveling times, producing a well-defined minimum on each case.
We define with a marker the fleet sizes at the minimum average journey duration as optimum fleets.
Note that higher willingnesses to share lowered the average journey duration and the fleet sizes in the optima.
The line ``direct travel'' indicates the in-vehicle travel time for immediate service, excluding waiting, detour, and abandonment penalty.
It represents the aggregated congestion model of the city, based on the MFD of Fig. \ref{fig:plot_simu_data}D, and it emphasizes the importance of integrating a congestion model in the analysis.
\href{https://drive.google.com/file/d/1ip5aMyco0xDX6xA176q-CAcMSD5JXL0X/view?usp=sharing}{Movie S1} shows how the identification of the optimum fleets relates to the number of idle vehicles at peak-hour getting close to zero.

The parking strategy helps to control congestion and to avoid unnecessary vehicle presence in the streets.
The average journey duration and waiting times reach a minimum, and they do not rise for larger fleet sizes (Fig. \ref{fig:plot_wait_trip}B and \ref{fig:plot_wait_trip}D).
For instance, traveling speeds were 14\% larger, and the average journey duration was 1.4 minutes shorter.
However, with fewer vehicles cruising, distances to pick-up passengers increased, and minimum waiting times became 1 minute larger, on average.

\begin{figure}[!ht]
    \centering
    \includegraphics{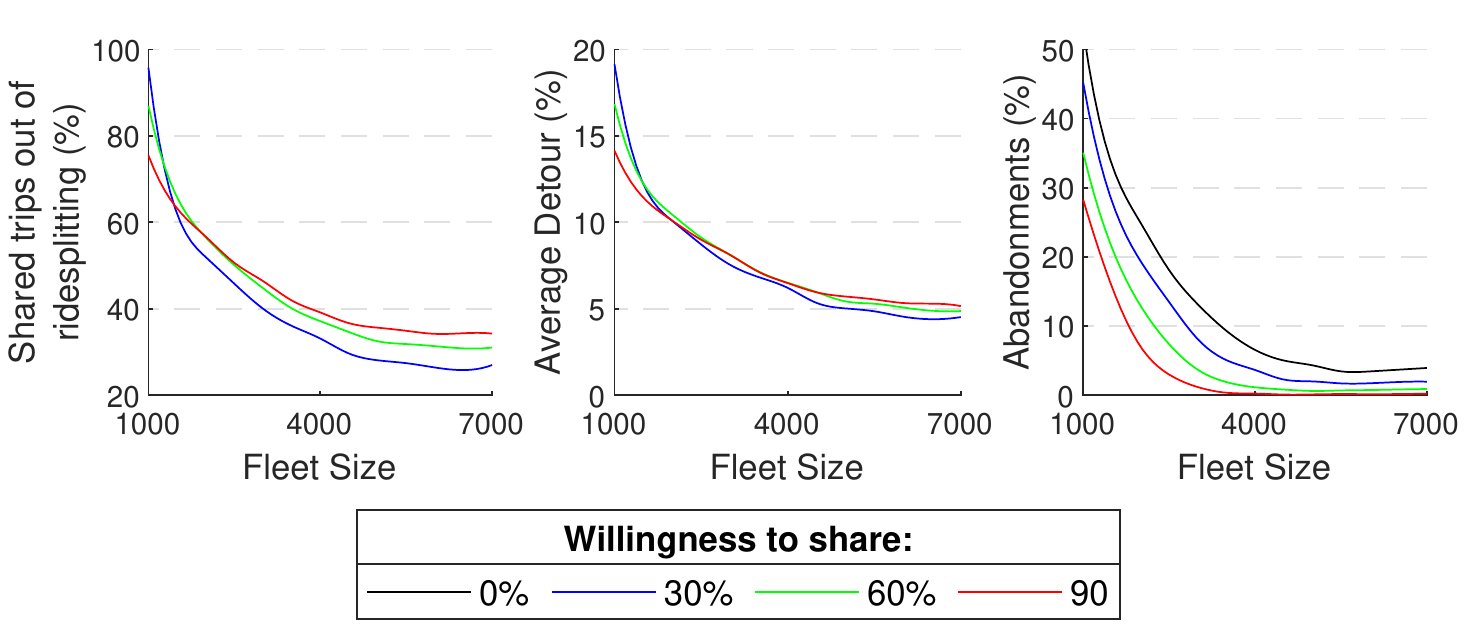}
    \caption{Shared trips fraction (accounting only for ridesplitting hired trips) and average detours for increasing fleets.}
    \label{fig:plot_shared_detour}
\end{figure}

Ridesplitting has two major uncertainties when trying to get more customers.
On the one side is the uncertainty of matching passengers, while on the other side is the extra time and distance that they will deviate from their initially designed trip.
In Fig. \ref{fig:plot_shared_detour} we explore these concerns in a growing fleet size perspective.
Although small fleets may provide higher chances for matching passengers, they face long detours and waiting times that increase abandonments.
On the other hand, large fleets provide the opposite situation.
In general, passengers' willingness to share has minor influence in the detours they face and their chances of matching another passenger.
All the previous indicates that fleet size is vital in identifying potential matches.

\begin{figure}[!ht]
    \centering
    \includegraphics[width=\textwidth]{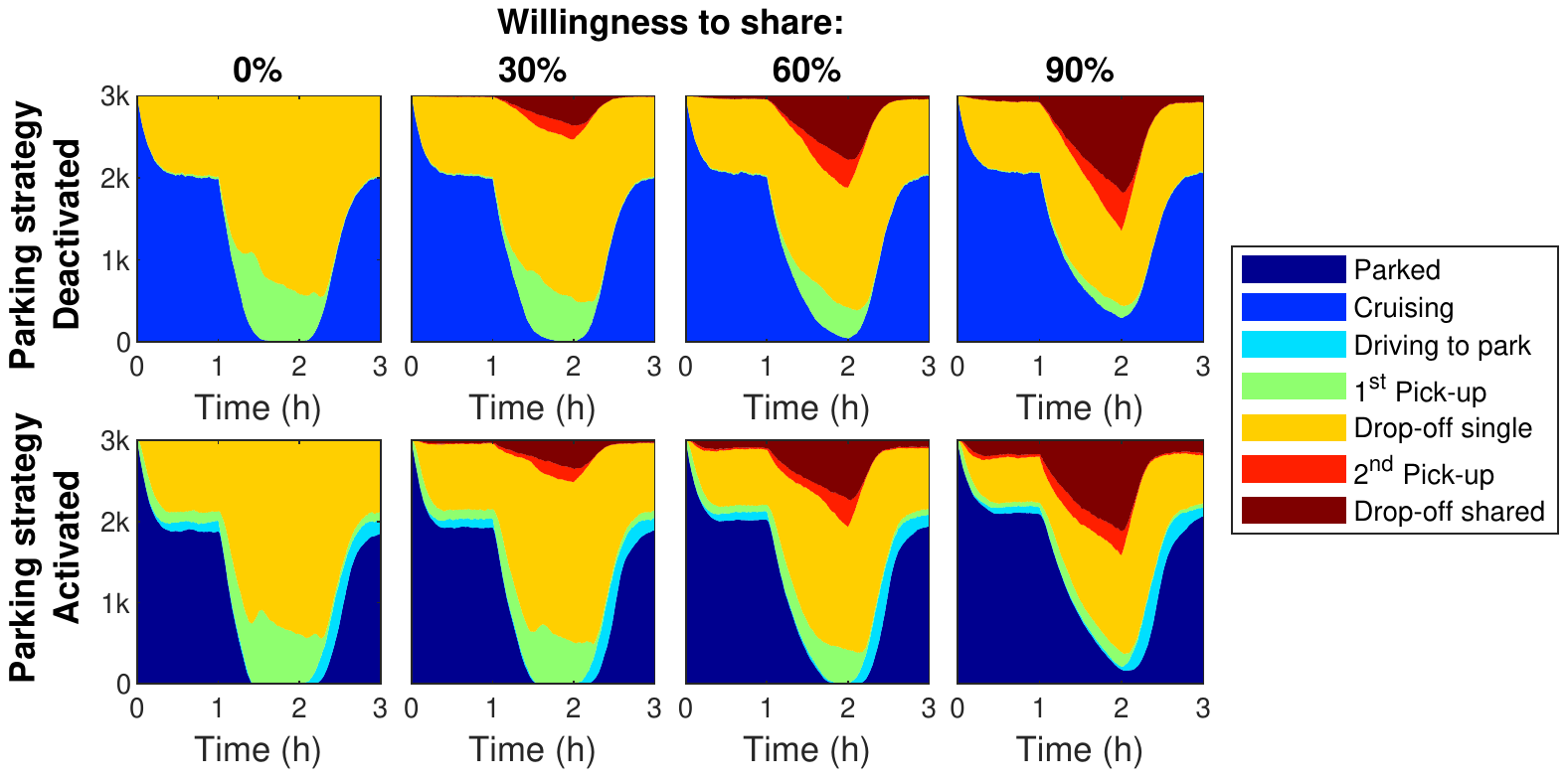}
    \caption{Number of vehicles in each state for instances with a fleet size of 3000 ride-sourcing vehicles, and varying willingness to share and use of the parking strategy.}
    \label{fig:plot_states}
\end{figure}

To understand these outcomes, we need a more extensive examination of these instances instances.
With a fixed fleet size, we can observe the situation of ride-sourcing vehicles through time and evaluate how it may influence service performance.
Fig. \ref{fig:plot_states} illustrates the system conditions through the number of vehicles in each state.
One can readily note the effect of the peak-hour over the system.
Firstly, idle vehicles rapidly become busy.
Secondly, in their absence, the number of vehicles working with shared trips (`2\textsuperscript{nd} pick-up' and `Drop-off shared' states) rises substantially.
The parking strategy does not allow empty vehicles to cruise, forcing them to park whenever idle.
Hence, idle vehicles may be in `parked' or `driving to park' states.
The use of the parking strategy enabled shared rides, even before the peak-hour (see `Drop-off shared' state in scenarios with 90\% willingness to share).
Table \ref{tab:vht_states} reinforces the previous, which shows more VHT for `Drop-off shared' states.
However, the VHT used to pick-up shared rides is smaller, especially for larger willingness to share.
In general, VHTs for busy vehicles do not change in order, but they fall more than ten times when observing empty vehicles (`Cruising' and `Driving to park' states).

\begin{table}[!ht]
\centering
\caption{Vehicle Hours Traveled (VHT) for private vehicles and for each state of ride-sourcing vehicles (fleet size of 3000 vehicles).}\label{tab:vht_states}
\begingroup
\setlength{\tabcolsep}{0pt}
\begin{tabular}{p{0.1102\linewidth}
P{0.1394\linewidth}
P{0.1102\linewidth}
P{0.1067\linewidth}
P{0.1067\linewidth}
P{0.1067\linewidth}
P{0.1067\linewidth}
P{0.1067\linewidth}
P{0.1067\linewidth}
}
\hline
\multirow{3}{0.1102\linewidth}{Parking strategy} &
\multirow{3}{0.1394\linewidth}{Willingness $\mbox{to share}$} &
\multirow{3}{0.1102\linewidth}{Private vehicles} &
\multicolumn{6}{c}{VHT per state} \\ \cline{4-9}
& & & Cruising & Driving to park & 1\textsuperscript{st} pick-up & Drop-off single & 2\textsuperscript{nd} pick-up & Drop-off shared \\ [2pt] \hline
\multirow{5}{*}{   \rotatebox{90}{Deactivated}} & 0\% & 33574 & 4860 & 0 & 792 & 0 & 4687 & 0 \\ [2pt]
 & 30\% & 32013 & 4903 & 0 & 617 & 126 & 4284 & 316 \\ [2pt]
 & 60\% & 30974 & 5205 & 0 & 391 & 221 & 3761 & 723 \\ [2pt]
 & 90\% & 30616 & 5584 & 0 & 157 & 262 & 3208 & 1106 \\ [2pt] \hline
\multirow{5}{*}{   \rotatebox{90}{Activated}} & 0\% & 31504 & 0 & 428 & 1020 & 0 & 4477 & 0 \\ [2pt]
 & 30\% & 29968 & 0 & 430 & 850 & 134 & 3997 & 350 \\ [2pt]
 & 60\% & 28210 & 0 & 432 & 594 & 221 & 3349 & 783 \\ [2pt]
 & 90\% & 27220 & 0 & 420 & 327 & 283 & 2504 & 1255 \\ [2pt]
\hline
\end{tabular}
\endgroup
\end{table}

\begin{figure}[!ht]
    \centering
    \includegraphics[width=\textwidth]{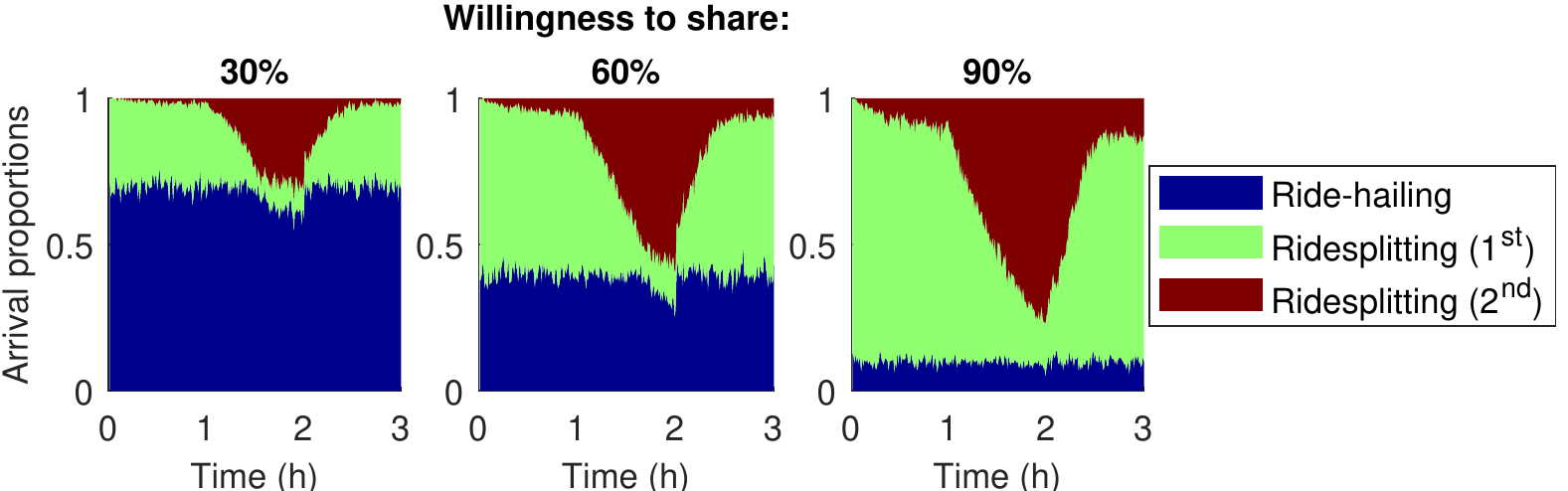}
    \caption{Timely computed arrival proportions for both services (ride-hailing and ridesplitting) and different willingness to share in a scenario with a fleet of 3000 vehicles.}
    \label{fig:plot_arrival_rates}
\end{figure}

System congestion may produce changes in the participation of each service (ride-hailing and ridesplitting) in the number of served passengers.
Fig. \ref{fig:plot_arrival_rates} illustrates the allocation of passengers between each service, highlighting shared rides of ridesplitting for scenarios with a fleet size of 3000 vehicles.
Note that we did not plot the results for willingness to share of 0\% because all passengers hire ride-hailing services (with no matching).
Mainly, ride-hailing demand remains constant for most of the simulation time.
The exception is the peak-hour, where ride-hailing loses space for ridesplitting beyond the willingness to share.
The reduction in ride-hailing trips occurs between 1.5h and 2h when demand is high, and the number of idle vehicles is small (see Fig. \ref{fig:plot_states}).
At the same time, only shared rides become available, another reason for increasing their proportions.
However, the shared trip proportion does not increase instantly with the increase in demand.
The later indicates that shared trips require a pool of vehicles with a single passenger to form before starting to share.

\subsection{Traffic and TNCs relation}

\begin{figure}[!ht]
    \centering
    \includegraphics[width=\textwidth]{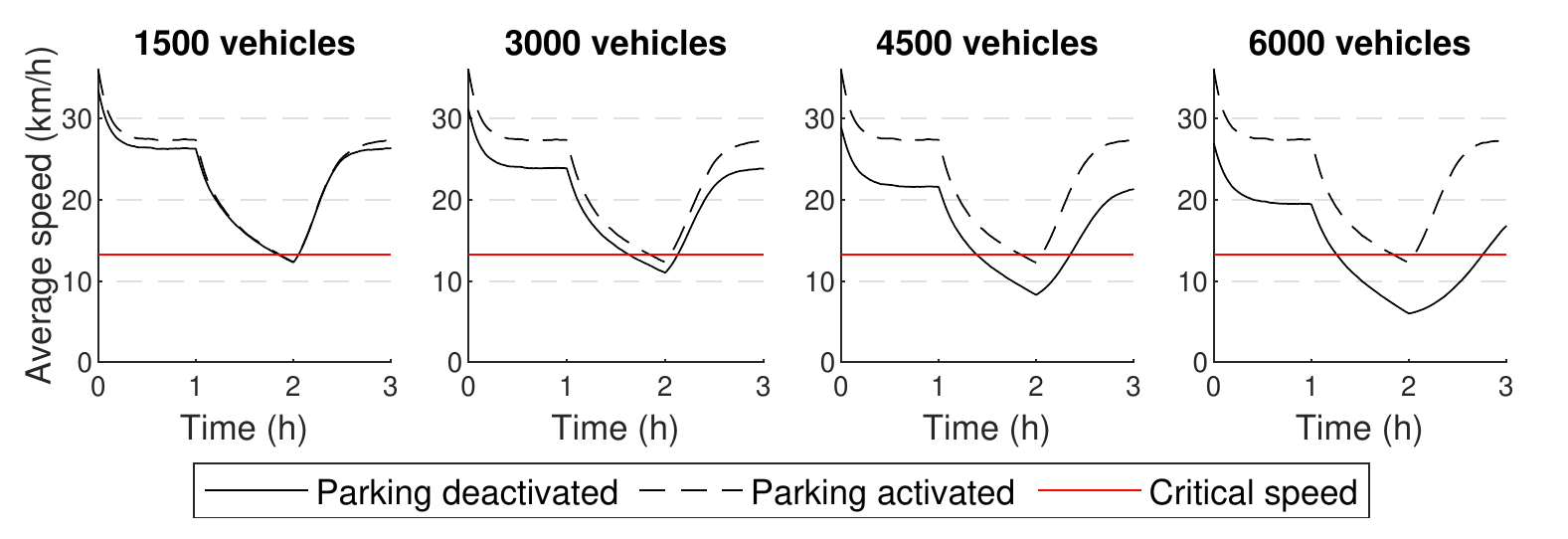}
    \caption{Growing fleets deteriorate average speeds and their restoration after the peak-hour. Parking idle vehicles have enhances average speeds independently of the fleet size.}
    \label{fig:plot_speeds}
\end{figure}

Ride-sourcing vehicles compose urban traffic, influencing traffic performance depending on their actions.
Fig. \ref{fig:plot_speeds} reveals that traveling speeds take longer times to recover from the peak-hour for larger fleet sizes.
Parking idle ride-sourcing vehicles enhanced the recovery speed from the peak-hour, therefore, increasing the overall resilience of the system \citep{zhang_etal_2019}.
Under a fleet size of 3000 ride-sourcing vehicles, instances without the parking strategy reached the critical speed (speed which maximizes flow in the MFD) after 38 minutes after the start of the peak-hour and entered a hyper-congested state for 29 minutes.
On the other hand, with the same fleet size with the parking strategy active, they reached the critical speed after 51 minutes after the beginning of the peak-hour and entered a hyper-congested state for 12 minutes.
The previous numbers have only marginal changes when the parking strategy is active but worsen for larger fleets when it is inactive (Fig. \ref{fig:plot_speeds}).

\begin{figure}[!ht]
    \centering
    \includegraphics{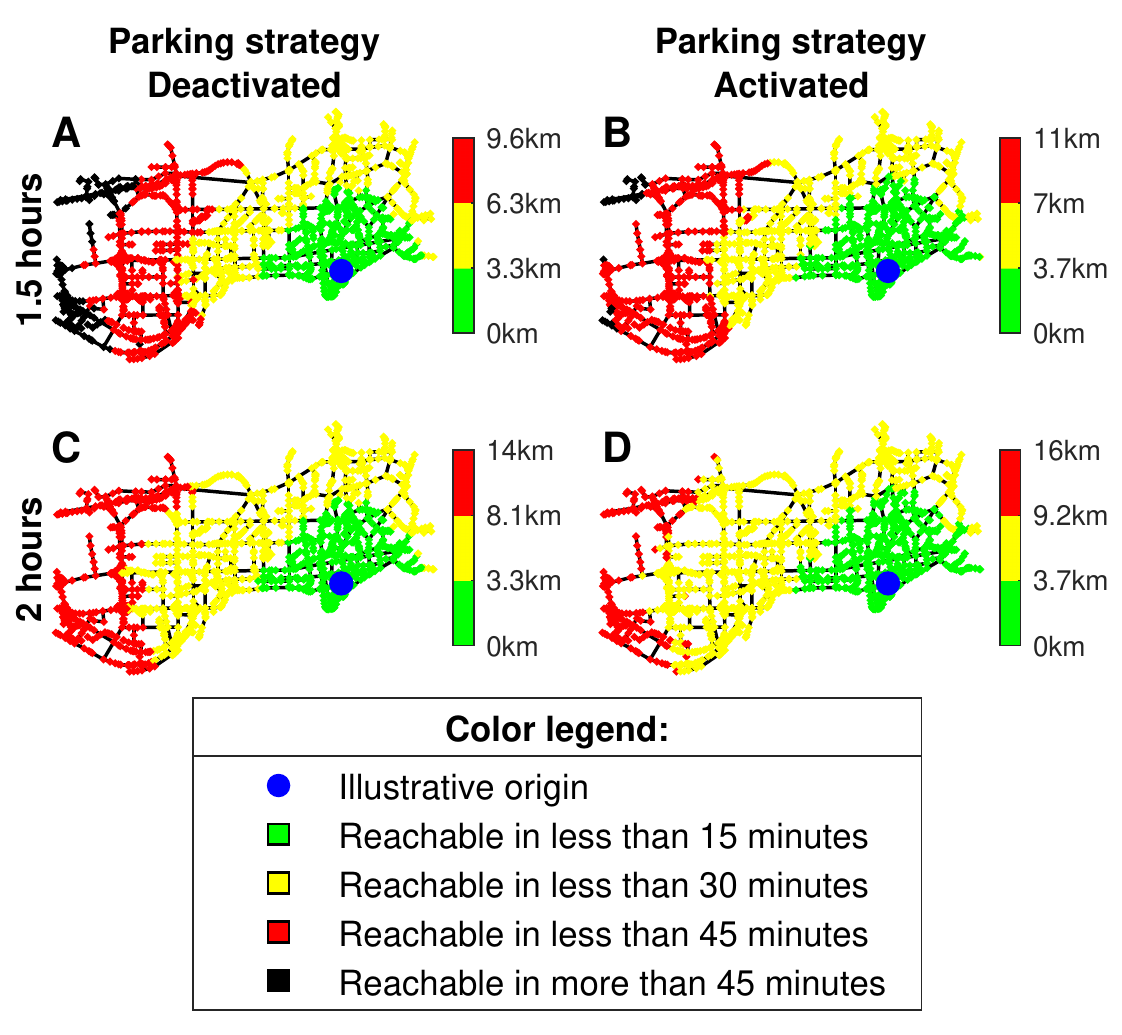}
    \caption{Taking idle vehicles from the streets enhances the area a traveler can reach in a certain time window and enhances the resilience of the system, in general. Dynamic reachable area in the modeled network (scenarios with 2500 ride-sourcing vehicles). Starting from an intersection in the central business district of Shenzhen (marked as a blue circle), the reachable area (and distance) that one can access within a certain time window (i.e., 15 min, 30 min, and 45 min) at certain simulated times (A and B 1.5 hours after simulation starts, (C and D) 2 hours after simulation starts). See \href{https://drive.google.com/file/d/1kJyzDUWMU21_ccYAL_tRIHklMGJtOiSy/view?usp=sharing}{Movie S2} for a complete observation of reachable areas over time.}
    \label{fig:plot_reach}
\end{figure}

Next, we explore in Fig. \ref{fig:plot_reach} the reachable area as a function of time starting from a node in the central business district.
As time advances, the driver can travel longer distances until reaching the whole network modeled.
A 0.5-hour difference in departure time changes significantly in the reachable area.
For example, departing 1.5 hours after the simulation start, a driver can reach a distance of around 6.3 kilometers, comprising 59\% of the simulated network, in 30 minutes, and this reachable area will extend to 74\% if departing 0.5 hours later.
However, traffic conditions recover faster when using the parking strategy, allowing a driver to travel 2.2 kilometers more by departing 0.5 hours later, whereas only 1.8 kilometers more without the parking strategy.
\href{https://drive.google.com/file/d/1kJyzDUWMU21_ccYAL_tRIHklMGJtOiSy/view?usp=sharing}{Movie S2} shows the evolution of reachable areas and speeds in the simulation.

\begin{figure}[!ht]
    \centering
    \includegraphics[width=\textwidth]{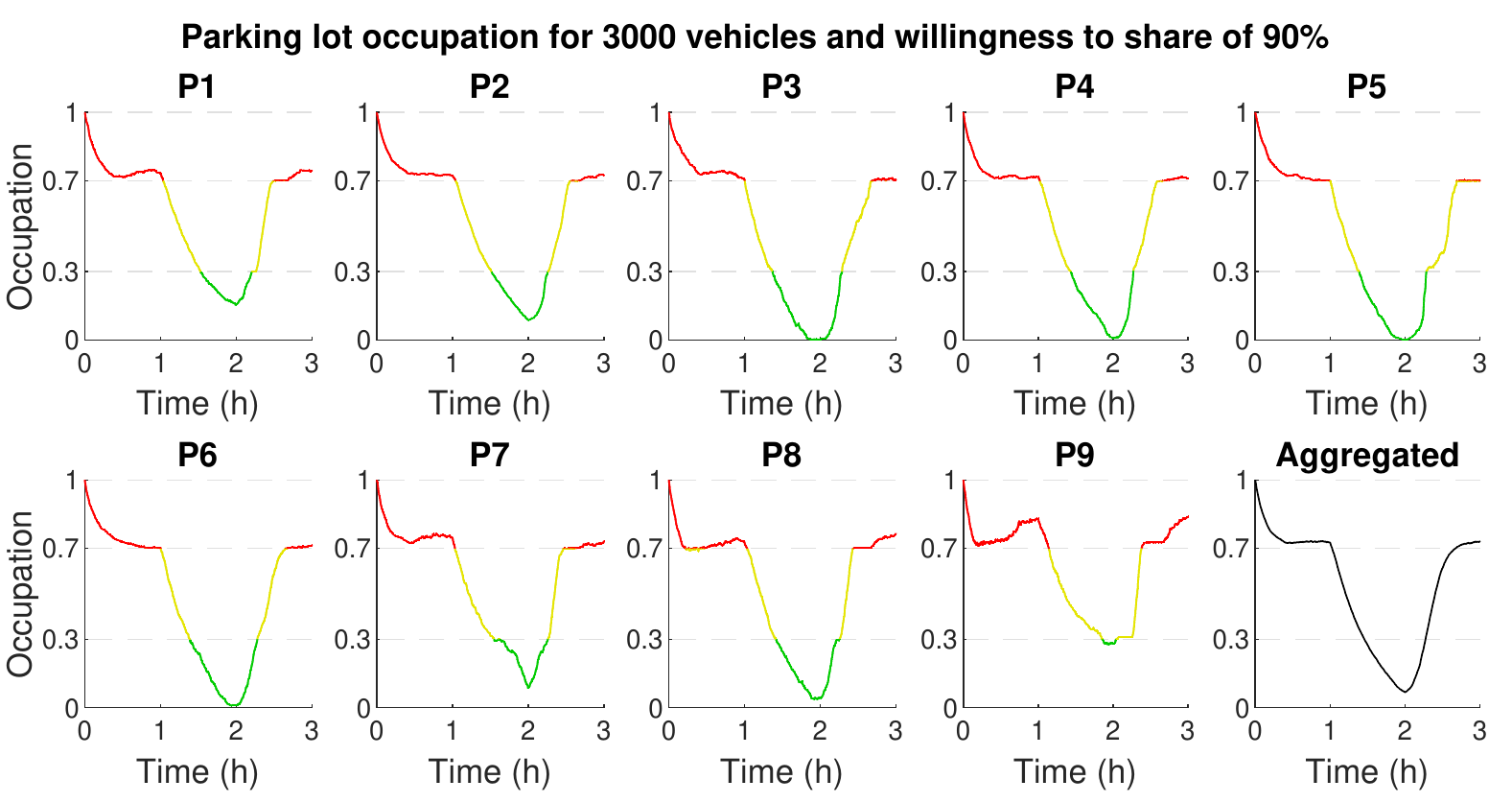}
    \caption{Parking lot occupation level and their instantaneous color flags.}
    \label{fig:plot_park_use}
\end{figure}

Fig. \ref{fig:plot_park_use} shows the number of assigned vehicles to each parking lot.
Firstly, parking lots empty similarly until the point they reach a green flag.
At peak-hour, all parking lots reach the green flag, and most of then have no vehicles around 2 hours of simulation.
Except for parking lot P9, which kept nearly 30\% of their capacity, very few vehicles remained parked.
Yet, this strategy could yield positive results for traffic conditions (Figs. \ref{fig:plot_speeds} and \ref{fig:plot_reach}).
The period after peak-hour illustrates the effects of the parking assignment algorithm (Algorithm \ref{alg:park}), where those parking lots that reached the flag thresholds have a lag before receiving new assignments.

\begin{figure}[!ht]
    \centering
    \includegraphics{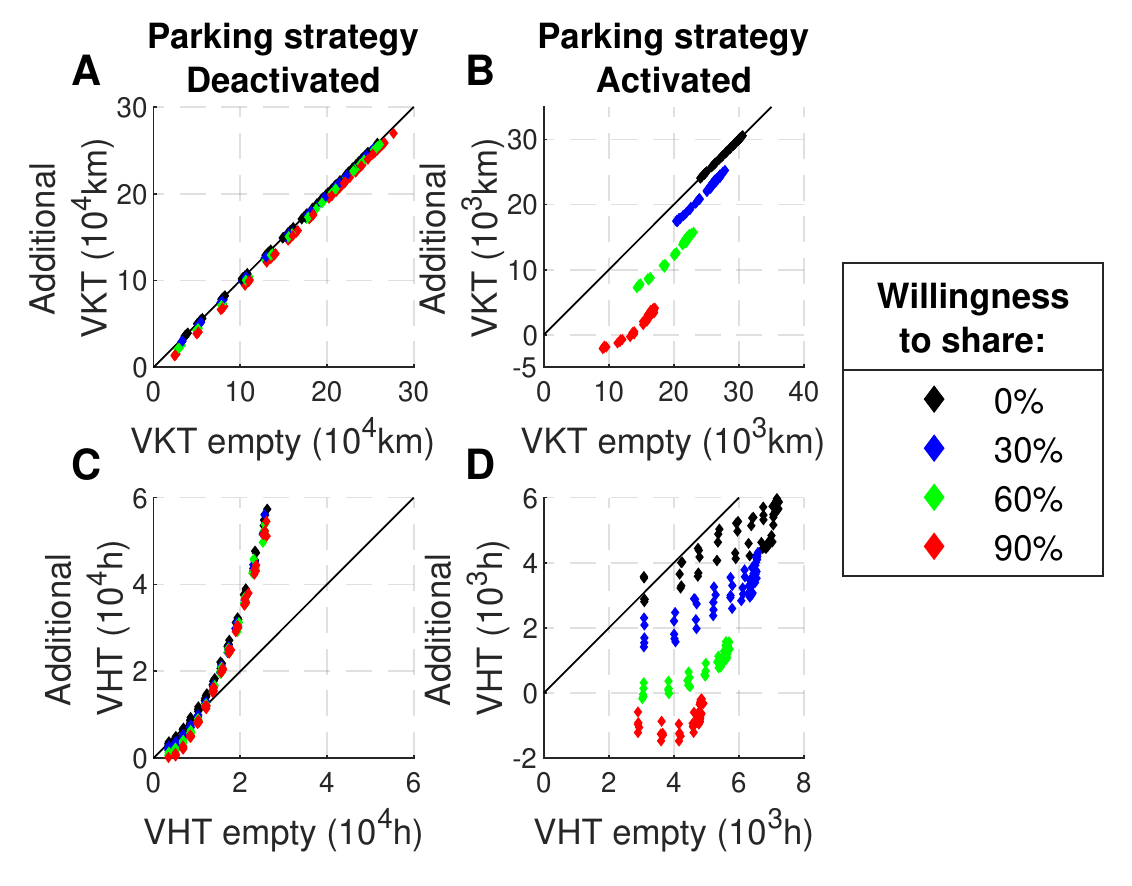}
    \caption{Increasing ride-sourcing fleets had higher impact on generating additional kilometers traveled (most of these traveled empty when searching for new passengers) than the will of passengers to share their rides. (A and B) Relationship between Vehicle Kilometers Traveled (VKT) generated by empty ride-sourcing vehicles and the additional VKT to direct travels. (C and D) Relationship between Vehicle Hours Traveled (VHT) of empty ride-sourcing vehicles and the additional VHT to a situation without ride-sourcing. (A) Additional VKT \textit{vs} VKT empty for the scenario without parking strategy. (B) Additional VKT \textit{vs} VKT empty for the scenario with parking strategy. (C) Additional VHT \textit{vs} VHT empty for the scenario without parking strategy. (D) Additional VHT \textit{vs} VHT empty for the scenario with parking strategy.}
\label{fig:plot_vkt_vht}
\end{figure}

VKT is a fundamental measure for transportation systems since it is related to the emission of greenhouse gases, such as CO$_2$ and other pollutants.
Moreover, it is associated with worse congestion, fuel consumption, and safety issues.
Ride-sourcing vehicles generate VKT not only when transporting passengers, but also when they are searching for them, resembling taxis.
Hence, we explore through Fig. \ref{fig:plot_vkt_vht} how ride-sourcing fleets generate additional VKT to the city.
Therefore, in a scenario with enlarging fleets of ride-sourcing, the only alternative to curb the growth of VKT is to take idle ride-sourcing vehicles from the streets.
We show on Eq. [\ref{Eq:add_VKT_i}] how ride-sourcing generated additional VKT ($\mbox{VKT}^{+}$) compared to a scenario where all travelers would travel alone to their destinations without the service (assuming no issues with parking).

\begin{align}
    {\mbox{VKT}^{+}} = \underbrace{\int^{t_f}_{t_i} { ( N^{N\!P}_{\mbox{\footnotesize{RSV}}} (t) N_{\mbox{\footnotesize{PV}}} (t) ) v (t) \, dt}}_{
    \begin{matrix}
    \mbox{VKT of all vehicles} \\
    \mbox{vehicles}
    \end{matrix}
    } - \underbrace{\sum_{\mbox{\footnotesize{TP}}}{p(\mbox{TP}^o_j , \mbox{TP}^d_j)}}_{
    \begin{matrix}
    \mbox{Sum of passengers'} \\
    \mbox{shortest paths}
    \end{matrix}
    } - \underbrace{\sum_{\mbox{\footnotesize{PV}}}{p(\mbox{PV}^o_i , \mbox{PV}^d_i)}}_{
    \begin{matrix}
    \mbox{Sum of private vehicles'} \\
    \mbox{shortest paths}
    \end{matrix}
    } \label{Eq:add_VKT_i}
\end{align}


\noindent
Here $t_i$ and $t_f$ indicate the beginning and the end of a simulated instance.
$N^{N\!P}_{\mbox{\footnotesize{RSV}}}(t)$ refers to the number of ride-sourcing vehicles on the streets (not parked), $N_{\mbox{\footnotesize{PV}}}$ refers to the number of private vehicles on the streets, and $v(t)$ refers to the speed on the network at time $t$.
$\mbox{p}(\cdot , \cdot)$ is the shortest path distance between two points in the network.
$TP$ is the group of all served passengers of the ride-sourcing service, and $\mbox{TP}^o_j$ and $\mbox{TP}^d_j$ are the origins and destinations of a passenger $j$, respectively.
$PV$ is the group of all travelers that used private cars (or taxis in case of abandonments), and $\mbox{PV}^o_i$ and $\mbox{PV}^d_i$ are the origins and destinations of a traveler $i$ from this group, respectively.
Remember that, private vehicles are assumed to use the shortest path, and to leave the system once reaching their destinations, thus Eq. [\ref{Eq:pv_equiv}] holds.

\begin{align}
    \int^{t_f}_{t_i} { ( N_{\mbox{\footnotesize{PV}}} (t) ) v (t) \, dt} = \sum_{\mbox{\footnotesize{PV}}}{p(\mbox{PV}^o_i , \mbox{PV}^d_i)} \label{Eq:pv_equiv}
\end{align}

\noindent
Thus, Eq. [\ref{Eq:add_VKT_i}] may be simplified to Eq. [\ref{Eq:add_VKT_f}].

\begin{align}
    {\mbox{VKT}^{+}} = \int^{t_f}_{t_i} { N^{N\!P}_{\mbox{\footnotesize{RSV}}} (t) v (t) \, dt}
    - \sum_{\mbox{\footnotesize{TP}}}{p(\mbox{TP}^o_j , \mbox{TP}^d_j)} \label{Eq:add_VKT_f}
\end{align}

Fig. \ref{fig:plot_vkt_vht}A shows that the VKT generated from empty vehicles has a high positive correlation with additional VKT independently of the willingness to share in instances where vehicles are allowed to cruise for passengers.
On the other hand, fleet size seems to be central for both empty and additional VKT.
However, Fig. \ref{fig:plot_vkt_vht}B shows that taking empty ride-sourcing vehicles from the streets (through an parking strategy) decreases the additional VKT almost ten times (note different units between Figs. \ref{fig:plot_vkt_vht}A and \ref{fig:plot_vkt_vht}B).
Moreover, willingness to share decreases and limits additional VKT.
Note that the only cases where ride-sourcing could decrease VKT (negative values of Additional VKT) occurred when fleets were small (smaller than the optima seen in Fig. \ref{fig:plot_vkt_vht}), willingness to share was high, and the elegant parking strategy is active.

On Fig. \ref{fig:plot_vkt_vht}C, Vehicle Hours Traveled (VHT) growth indicates that larger ride-sourcing fleets deteriorate traffic conditions for all users, reducing travel speeds.
There is a decrease in the order of additional VHT with the control over cruising ride-sourcing vehicles (Fig. \ref{fig:plot_vkt_vht}D).
However, in Fig \ref{fig:plot_vkt_vht}D, VHT decreased with a higher willingness to share because of higher speeds after a peak on demand.

\subsection{Revenues and Price of Anarchy}

The previous section showed that while larger fleet sizes decrease waiting time for passengers, this creates a higher congestion level and lower quality of service for all private modes of transport.
It is known that in competitive markets where different jurisdictions have not the same objective function, the system can reach states far from optimal welfare (see, for example, \citet{lamotte_etal_2017}).
While our work does not analyze equilibrium conditions between competitive players, we intend to show that TNC's decisions concerning fleet sizes can create problematic states for network congestion and even TNC driver profit.

\begin{figure}[!ht]
    \centering
    \includegraphics{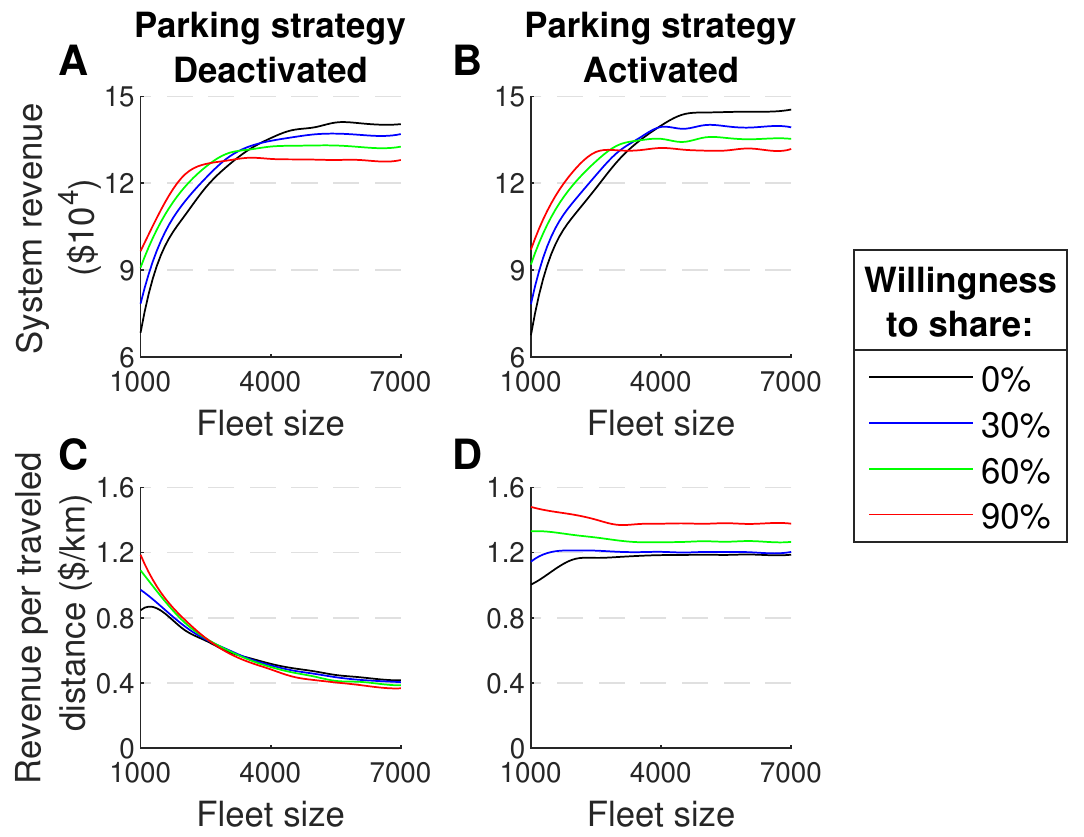}
    \caption{Higher willingness to share has higher revenues to smaller fleets but lower revenues for large fleets. (A and B) Total revenue of the system, increasing with fleet sizes. (C) Revenue per traveled kilometer of the ride-sourcing service, deteriorated by empty traveled kilometers. (D) Revenue per traveled kilometer of ride-sourcing service, preserved from deterioration.}
    \label{fig:plot_revenues}
\end{figure}

We also confirm that, under constant fares, it is attractive for the TNCs that the fleets grow.
Both Figs. \ref{fig:plot_revenues}A and \ref{fig:plot_revenues}B show increasing revenues for growing fleet sizes until a maximum point, which does not decrease afterward.
The interesting point here is that a higher willingness to share reduces the maximum revenue, but it also decreases the necessary fleet size to get such a result.
When willingness to share is high (60\%, or 90\%), plateaus of maximum revenues start close to optimum fleet sizes (Figs. \ref{fig:plot_wait_trip}C and \ref{fig:plot_wait_trip}D).
On the other hand, a lower willingness to share urges larger fleet sizes.
The broader adoption of sharing generates more revenue for the system and the driver, for adequately sized fleet sizes.
Nonetheless, lower willingness to share reaches higher revenue for larger fleets.
Two points cast light upon this.
Firstly, the costs are higher for passengers that do not want to share their rides (US\$2.20 \textit{vs} US\$2.00 as fixed cost and an additional of US\$1.00 \textit{vs} US\$0.80 per kilometer, respectively).
Secondly, revenue is lower for small fleets because of abandonments, which decrease as the fleet grows, and ride-sourcing vehicles get closer to the passengers (see Figs. \ref{fig:plot_wait_trip}A and \ref{fig:plot_wait_trip}B).
It is worth mentioning that Figs. \ref{fig:plot_revenues}C and \ref{fig:plot_revenues}D show that the empty VKT of ride-sourcing drivers deteriorate their revenue per kilometer traveled, a proxy for their earnings.
In both scenarios, drivers would prefer to drive only for single passengers, when the passenger pool for sharing is small (low willingness to share), and both proxies point to this, with an exception for small fleet sizes.
These results consider a fixed number of drivers and no surge pricing policy.

We can also approximate the price of anarchy the ride-sourcing systems can create in transportation networks.
We consider, as a system optimum solution, the situation that minimizes total system delay (red marker in fig. \ref{fig:plot_wait_trip}C), a fleet size of 2500 vehicles, willingness to share is 90\%, and the average journey time of 15.6 minutes.
We also consider a selfish optimum as the one that maximizes the revenue of TNCs, thus, where there is no willingness to share (nowadays, ridesplitting trips in TNCs correspond to about 5\% \cite{li_etal_2019}), based on fig. \ref{fig:plot_revenues}A with a fleet size of 5500 vehicles.
In this case, the PoA is 1.37, whereas with a willingness to share of 30\% and a restricted fleet size that minimizes system delays (blue point in fig. \ref{fig:plot_wait_trip}C it decreases around to 1.13 creating 11.5\% savings in travel times.
Note that when the number of cruising vehicles decreases through parking strategies, the PoA remains small as the optimal fleet size to minimize delays is close to the fleet size that maximizes revenue for TNC company (compare figs. \ref{fig:plot_revenues} D with \ref{fig:plot_wait_trip}B).
In the worst case with the parking strategy, PoA is 1.07.

\citet{tirachini_lobo_2019} analyzed for static conditions without congestion how incentives of the company and pricing strategies can influence the number of drivers that register for TNC services.
Analyzing this type of equilibrium game for similar settings as our problem (with congestion dynamics, empty kilometer traveled) can reach additional interesting insights.
This should be a research priority.

\section{Final considerations}

In this paper, we investigated the effect of expanding fleet sizes for TNCs, passengers with different willingness to share, and operational strategies over congestion conditions.
We highlight that, by omitting dynamics of congestion in rides-sourcing studies to focus on matching strategies or rebalancing vehicles in static environments, different conclusions with possibly unrealistic performance measures are obtained.

Results show that sharing (allowing ridesplitting with a large pool of passengers) by itself is not capable of decreasing the system's VKT if there is no control over the fleet (its size and operation).
To reduce emissions (by reducing VKT), TNCs should change their \textit{modus operandi}; in a way to avoid that their fleet cruises without an assigned passenger.
On the other hand, sharing decreases the number of vehicles needed to maximize coverage and minimize service times.
Furthermore, in case it is not possible to avoid TNCs' fleets cruising for passengers, increases in the willingness to share can minimize both waiting times and service times.
For adequately sized fleet sizes, the adoption of sharing is related to higher revenues (for the system and the driver).
Therefore, so that ride-sourcing becomes a sustainable service, it must change its operations to remove vehicles without passengers from the streets, and passengers must become more receptive to ridesplitting at the same time.
However, it is outside the scope of this paper to combine different transportation modes (public ones, in particular) and surge pricing policies.
Therefore, we do not cope with the detailed modeling of decisions on the mode choice for travelers.
This should be a relevant further research priority.

\bibliographystyle{cas-model2-names}

\bibliography{cas-refs}

\end{document}